\documentclass[preprint2]{aastex}
\def\lsim{\lower.5ex\hbox{$\; \buildrel < \over \sim \;$}}
\def\gsim{\lower.5ex\hbox{$\; \buildrel > \over \sim \;$}}

\def\t{\ifmmode {\tau} \else $\tau$ \fi}

\def\ref{\noindent \hangafter=1 \hangindent=0.7 truecm}

\def\cm{\ifmmode {\rm cm}^{-1} \else cm$^{-1}$ \fi}
\def\s{\ifmmode {\rm s}^{-1} \else s$^{-1}$ \fi}
\def\cc{\ifmmode {\rm cm}^{-3} \else cm$^{-3}$ \fi}
\def\cs{\ifmmode {\rm cm}^{-2} \else cm$^{-2}$ \fi}
\def\g{\ifmmode \gamma \else $\gamma$\fi}
\def\G{\ifmmode \Gamma \else $\Gamma$\fi}

\def\kms{\ifmmode {\rm km\ s}^{-1} \else km s$^{-1}$\fi}

\singlespace
\begin{document}

\title{The Optical-Near-IR Spectrum of the M87 Jet From HST Observations}

\author{Eric S. Perlman \altaffilmark{1,2,3}, John A. Biretta\altaffilmark{3},
William B. Sparks\altaffilmark{2,3}, F. Duccio Macchetto\altaffilmark{3,4}, \and J. Patrick Leahy \altaffilmark{3,5}}

\altaffiltext{1}{Department of Physics, Joint Center for Astrophysics,
University of Maryland-Baltimore County, 1000 Hilltop Circle, Baltimore, MD
21250}

\altaffiltext{2}{Department of Physics and Astronomy, Johns Hopkins University, 3400 North Charles Street, Baltimore, MD 21218}

\altaffiltext{3}{Space Telescope Science Institute, 3700 San Martin Drive, Baltimore,MD 21218, USA}

\altaffiltext{4}{Affiliated with the Astrophysics Division of the European Space Agengy, ESTEC, Noordwijk, Netherlands}

\altaffiltext{5}{University of Manchester, Jodrell Bank Observatory, Macclesfield, Cheshire, SK11  9DL, UK}

\email{perlman@jca.umbc.edu}

\begin{abstract}

We present 1998 HST observations of M87 which yield the first single-epoch
optical and radio-optical spectral index image of the jet at 0.15''
resolution.  We find $\langle \alpha_{ro} \rangle \approx 0.67$, comparable to
previous measurements, and $\langle \alpha_o \rangle \approx 0.9$ ($F_\nu
\propto \nu^{-\alpha}$), slightly flatter  than previous workers.  Reasons for
this discrepancy are discussed. These observations reveal a large variety of
spectral slopes.  Bright knots exhibit significantly flatter spectra than
interknot regions.  The flattest spectra ($\alpha_o \sim 0.5-0.6$; comparable
to or flatter than $\alpha_{ro}$) are found in the two inner jet knots (D-East
and HST-1) which contain the fastest superluminal components.  The flux maximum
regions of other knots have $\alpha_o \sim 0.7 - 0.9$.

The maps of $\alpha_o$ and $\alpha_{ro}$ appear poorly correlated. In knots A,
B and C, $\alpha_o$ and $\alpha_{ro}$ are  essentially anti-correlated with one
another.  Near the flux maxima of two inner jet knots (HST-1 and F), changes in
$\alpha_{ro}$ appear to lag changes in $\alpha_o$, but in two other knots (D
and E), the opposite relationship is observed.  This is further evidence that
the radio and optical emissions of the M87 jet come from substantially
different physical regions.   The delays observed in the inner jet are 
consistent with localized particle acceleration in the knots, with $t_{acc} <<
t_{cool}$ for optically emitting electrons  in knots HST-1 and F, and $t_{acc}
\sim t_{cool}$ for optically emitting electrons  in knots D and E.  Synchrotron
models fit to the radio-optical data yield $\nu_B \gsim 10^{16}$ Hz for knots
D, A and B, and somewhat lower values, $\nu_B \sim 10^{15}- 10^{16}$ Hz, in
other regions of the jet. If the X-ray emissions from knots A, B and D are
co-spatial with the optical and radio emission, we can strongly rule out the
``continuous injection'' model, which overpredicts the X-ray emissions by large
factors.  Because of the short lifetimes of X-ray synchrotron emitting
particles, the X-ray emission likely traces sites of particle acceleration and
fills volumes much smaller than the optical emission regions.

\end{abstract}

\section{Introduction}

The giant elliptical galaxy M87 hosts the best-known extragalactic synchrotron
jet.  As a result of its proximity (distance = 16 Mpc, Tonry 1991),
particularly high resolution studies of its structure are possible ($1'' = 78$
pc).  The synchrotron nature of its emissions were first demonstrated by Baade
(1956).

Many workers have observed the M87 jet in the radio, optical and X-rays (Perola
\& Tarenghi 1980; Stocke et al. 1981; Schreier et al. 1982; Smith et al. 1983;
Biretta, Owen \& Hardee 1983; Killeen et al. 1984; Keel et al. 1988;
Perez-Fournon et al. 1988; Owen, Hardee \& Cornwell 1989; Biretta, Stern \&
Harris 1991, hereafter BSH91; Meisenheimer, R\"oser \& Schl\"otelburg 1996,
hereafter MRS96; Sparks, Biretta \& Macchetto 1996, hereafter SBM96; Zhou
1998).  Those data have revealed complex structure, with several bright
radio-optical knots, as well as significant information on the broadband
spectrum of the jet.

The first HST observations of M87 (Boksenberg et al. 1992) revealed
considerable detail in the optical jet.  Its large scale radio and optical
structures are remarkably similar; differences only appear at $\sim 0.1''$
scales (SBM96).  More recent HST observations have revealed apparent
superluminal motion in several components at speeds up to 6$c$ (Biretta et al.
1999; Biretta, Sparks \& Macchetto 1999, hereafter BSM99), linking the
properties of M87 with those of BL Lacertae objects.  HST observations have
also served to illuminate the jet's axial structure for the first time, as the
radio and optical polarized structures on $0.2''$ scales show strong and
consistent differences which indicate that higher-energy particles inhabit
regions closer to the jet axis (Perlman et al. 1999, hereafter P99).

The first ground-based multiband, contemporaneous optical photometry of the jet
(BSH91) revealed that its spectrum is considerably steeper in the optical than
in the radio.  More recent ground-based work (MRS96) confirmed this picture,
but disagreed with BSH91 on the spectral index of various jet components. Those
workers also presented the first detailed fits of synchrotron emission models
to broadband, data produced both in their own observations and in those of
previous workers (references above) and predicted synchrotron break frequencies
$\sim 10^{14} - 10^{15}$ Hz.  Pre-COSTAR HST data (SBM96) revealed considerable
narrowing of the jet from the radio to the optical, and flatter radio-optical
spectra in bright knots.  Those workers also extracted optical spectra for
large regions of the jet and were the first to fit synchrotron emission models
to radio-optical data for the jet at HST resolutions.  The model fits indicated
stronger magnetic fields, and shorter particle lifetimes in bright knot
regions.

Despite this large volume of work, several important questions remain
open.  For example, the presence of optical synchrotron radiation
requires that electrons must be accelerated to at least $\gamma =
10^6$.  Yet the mechanism by which particles are accelerated is poorly
known, as are the dynamics of the synchrotron spectrum with time.
Moreover, the origin of the X-ray emission from the jet is still not
completely constrained (BSH91, Harris et al. 1997, Neumann et
al. 1997), and pre-COSTAR HST data were not sufficient to provide a
high signal to noise map of the jet's spectral index structure
(SBM96).  

Here we present 1998 HST observations of the M87 jet, spanning seven bands
between $\lambda =0.3-2.05$ microns.   These data represent the highest
resolution spectral data ever gathered on any extragalactic jet.  To analyze
the broadband continuum shape, we also include observations taken at 15 GHz
with the VLA in 1994.  These data allow us to resolve several important issues
left unanswered by previous works. 

In \S 2, we will discuss the observational setup and data reduction
procedures.  Our results will be presented in \S 3, including both
radio-optical and optical spectral index maps of the jet.  In \S 4 we compute
synchrotron spectrum fits using our radio-optical data plus X-ray data from the
literature. In \S 5 we compare our results with those of previous authors. In
\S 6 we discuss the physical implications of our results.

\section {Observations and Data Reduction}

\subsection{HST Observations: Design}

We imaged the M87 jet with the WFPC2 and NICMOS aboard HST on 25 and 26
February 1998.  Data were obtained through six filters, spanning the wavelength
range 0.3-1.6 microns.  A seventh near-IR band (2.05 microns) was also
scheduled for 26 February, but had to be reobserved on 4 April 1998, due to a
loss of guide-star lock.  In Table 1, we list important details of these
observations.

To maximize the observable portion of the jet, the February observations were
done with the HST oriented such that the jet fell along a chip diagonal. 
Unfortunately an equally good orientation was not available on 4 April; as a
result, the F205W data do not include the region within $1''$ of the nucleus. 
In the F110W and F160W bands, the small size of the NIC1 detector made it 
necessary to observe at two positions. The majority of the observing time for
those bands was used on the inner jet due to its lower surface brightness.  To
maximize resolution and minimize the effect of known instrumental problems such
as grot (Sosey \& Bergeron 1999), bad columns and warm pixels, observations in
the F205W, F160W, F110W and F814W bands were dithered. We used two-position
dithers in F205W, F160W and F110W, while for the F814W observation, we obtained
four images in a three-position pattern with sub-pixel dithering. In the F205W
band, two dithered 192s observations per orbit were done at a position $700''$
away to measure the thermal background.

\subsection{HST Observations: Data Reduction}

We reduced each dataset using the best recommended flatfields, biases, darks
and illumination correction images.  The IRAF task CRREJ was used to combine
CR-split WFPC2 images and eliminate cosmic rays.  For the F814W data, we used
CRREJ first on the 2 images taken at the central dither position, then shifted
the PC images taken at the 2 other positions to be consistent with one another
using IMLINTRAN, and then combined those with CRREJ.  The product images were
then combined using DRIZZLE (Fruchter \& Hook 1998, Gonzaga et al. 1998) onto a
grid with pixels  aligned 45$^\circ$ to the original orientation of the image
and smaller by a factor of $\sqrt 2$. DRIZZLE was also used for the other WFPC
images in order to correct for the geometric distortion in the PC chip
(Holtzmann et al. 1995), using  solutions by Casertano et al. (in prep.)

Unequal pedestal effects in the NICMOS data were eliminated with UNPEDESTAL
(van der Marel 1998).  No overall pedestal correction was computed, since the
signal from the galaxy overwhelms the pedestal. This resulted in only small
differences ($\sim 0.002$ ADU/pix/s) near the join of the two point positions
used in the F110W and F160W images.  In order to minimize color-dependent terms
in the NICMOS F110W and F160W flat-fields, we also used color-dependent
``superflats'' made from long observations of Io (Storrs, Bergeron \& Holfeltz
1999).  The color dependent terms are reduced significantly in the
longer-wavelength bands; moreover, the same concerns are also addressed by the
thermal background images taken for the F205W data. The NICMOS images were
combined using scripts from the DRIZZLE Handbook (Gonzaga et al. 1998) and
offsets computed from the {\it jitter} files using the script BEARING by E.
Bergeron (mentioned in Gonzaga et al.).  To correct for geometric distortion
and the slightly rectangular pixels, we used solutions by Cox et al. (1997). 
Thermal background images for the F205W band were combined in the same way and
then subtracted.

Galaxy subtraction was done using the tasks ELLIPSE, BMODEL and IMCALC. To
successfully model the galaxy, we masked out the jet and SW hotspot, as well as
all globular clusters and stars.  This is of necessity an iterative process, as
the fainter globular clusters do not become apparent until galaxy subtraction
is done.  Failure to mask a globular cluster would produce a circular
``ringing'' at the distance of the cluster.  In the WFPC2 images, we only fit a
galaxy model to the data on the PC chip, in order to minimize the effect of the
chip join, which was found to produce a circular ``ringing'' near the radius of
knot F in fits done with the entire 4-chip WFPC2 mosaic.  Due to the faintness
of the galaxy in the UV,we were only able to subtract the galaxy from the F300W
image after smoothing it with a Gaussian.  In the NICMOS mosaics, it was
necessary to manually mask out regions where no data exists.  The smallness of
the NIC1 chip only made it possible to model the galaxy in the F110W and F160W
images to a distance of $\sim 19''$ from the nucleus.  This also necessitated
an iterative approach to finding the correct ellipticity and PA of the model
galaxy in the F110W and F160W bands ({\it n.b.,} it was not possible to assume
a surface brightness model derived from the WFPC2 images and scaled to the
nuclear flux because of significant large-scale color gradients within the
galaxy itself).  Subsequent testing revealed that in all bands we were able to
achieve residuals no larger than a few ADU/pix, with no large-scale trends
which might indicate incorrect ellipticity or ``ringing'' in most regions,
although there are significant dust known lanes in the inner galaxy (Sparks et
al. 1993). The dust lanes affect only one region of the jet, knot HST-1.  RMS
errors from the galaxy subtraction are included in our error estimates in each
case by adding in quadrature.

Flux calibrated images were obtained by multiplying the reduced and
galaxy-subtracted image by using the task CALCPHOT in SYNPHOT.  We used
CALCPHOT to do this task rather than the information in the headers because it
allowed us to assume a spectral index value (we used $\alpha=0.9$, which is
close to the mean spectral index of the jet) and thus refine the flux
calibration{\footnote{note that the differences between the values derived for
$\alpha=0.9$ and the values in the headers were in all cases $<3\%$}} of the
PHOTFLAM value from SYNPHOT.  This also allowed us to apply a decontamination
correction (Baggett et al. 1997, Baggett \& Gonzaga 1998), which affected most
bands by $\lsim 3\%$, but had a larger effect on the F300W data (8\%). To
simulate the size of the aperture correction due to the extreme wings of the
PSF, we used the knot A+B+C region and a TinyTim PSF for the F205W+NIC2
combination.  Those simulations led us to adopt an aperture correction of 1.107
for regions which extend $\sim 1''$ (on each side) beyond regions of the jet
with {\it visible} flux (such simulations are recommended for estimating the
aperture correction for extended sources; see NICMOS team 1999).  Conversion
from $F_\lambda$ to $F_\nu$ was done at each band's pivot wavelength
($\lambda_{\rm eff}$ in Table 1).

Considerable effort was expended to estimate the uncertainties in these data. 
Several possible sources of error were accounted for, including zeropoint
errors in the SYNPHOT calibration (Colina et al. 1998 for NICMOS; Baggett \&
Gonzaga 1998, Baggett et al. 1997 and Whitmore \& Heyer 1995 for WFPC2;
typically this error is $\lsim 3\%$), possible errors due to assuming the wrong
slope for the  jet in SYNPHOT (at most 1-2\%, even if standard SYNPHOT
coefficients instead of ones calculated for $\alpha=0.9$ are used),
flatfielding errors ($\lsim 2\%$ for WFPC2, Whitmore \& Heyer 1997, but $\sim
3-5\%$ for NICMOS, Calzetti \& Gilmore 1998) and Poisson errors both in the
data and in the subtraction of the galaxy (see above).  These errors have been
propagated by adding in quadrature.

Once these steps were done, all images were rotated so that the jet was along
the x axis (the centerline of the jet was assumed to lie at PA 290.5$^\circ$),
using the positional and orientation information in the headers.  Images were
then convolved with Gaussians so that their resolution (FWHM) was equal to that
of the F205W image (TinyTim PSFs were not used because of the large number of
bands and difficulties associated with deconvolution tasks).  We then resampled
all images to the nominal NIC2 pixel size of $0.0758667''$/pix and placed them
on identical coordinate grids, with the nucleus at (20,200) on 400 $\times$ 400
pixel images.  Testing has revealed that the alignment of these images is
accurate to within 0.15 pixels.  The resolution of the spectral index maps we
show in \S\S 3 and 4 is therefore $\sim 0.15''$ FWHM, translating to 12 pc at
the distance of M87.

\subsection{VLA Data}

To analyze the jet's broadband spectrum, we include previously published radio
data (Zhou 1998, P99), obtained in February 1994 at 15 GHz.  The data reduction
procedures for this dataset have  been described in Zhou (1998) and P99.  To
facilitate direct comparison between the radio and optical images, the radio
image was resampled and shifted to the same frame of reference and pixel scale
as the HST images (see above).

While the radio data were obtained in February 1994, the possible effect of
variability is small.  Even for the largest radio variability found between
1993 and 1997 ($\sim 40\%$ for HST-1; Zhou 1998), the effect on $\alpha_{ro}$
is only $\Delta \alpha_{ro}=0.04$. The resolution of the radio and F205W images
is very similar ($0.14''$ FWHM compared to $0.15''$ FWHM) so we did not have to
convolve the radio image with a Gaussian before comparisons were done, and we
were able to use the same aperture correction.

\section{Radio-Optical and Optical Continuum Shape}

In Figure 1, we show the WFPC2 300W and F814W images, and the F160W NICMOS
image, all at full resolution.  The structures seen in the three bands are
similar, indicating that spectral curvature exists in various jet components
tend to be gradual. We examined the profiles on each HST image of slices
through resolved areas of the jet.  These comparisons were done on convolved
data only, to exclude resolution effects.  We confirm the result of SBM96 that
the jet is narrower in the optical than the radio.  However, based on our data
there is no evidence that significant narrowing takes place in the optical.  

\subsection{Spectral Index Fits}

We generated optical and radio-optical spectral index maps ($\alpha_o$ and
$\alpha_{ro}$; $F_\nu \propto \nu^{-\alpha}$), using simple weighted
least-squares fitting algorithms in IDL.  In Figure 2a, we show the optical
spectral index map.  Both the $\alpha_o$ and $\alpha_{ro}$ images were clipped
in regions where the signal to noise in either the F814W or radio image was
less than 4. In Figure 2b, we show the radio-optical spectral index
($\alpha_{ro}$) map.  The same figure also shows maps of synchrotron model fit
parameters in panels 2c and 2d (see \S 4 for description of those and
discussion).  The convolved F814W image is shown for comparison in Figure 2e.

The lack of ``ringing'' near compact features in the $\alpha_o$ map indicates
that the differing resolutions in the HST images were correctly accounted for,
and also that the registration is quite good.  There is a large steep-spectrum
region at the southern end of the A+B+C complex, but it is clearly real  since
(1) there is no corresponding feature near the southern end of the inner jet,
and (2)  there is more flux and extent to the south of the bright regions in
A+B+C than to the north (see the images in SBM96 and P99 as well as Figure 1).

In Figure 3, we show runs of $\alpha_o,~ \alpha_{ro}$, $\alpha_o-\alpha_{ro}$
and flux with distance along the axis of the jet in the inner jet ($0-10''$
from the nucleus).  Figure 4 shows the same plots for the outer jet ($10-20''$
from the nucleus).  The spectral indices given in those figures are
flux-weighted on a row-by-row basis, with flux summed in a $1.5''$ wide window
for Figure 3 and a $6''$ wide window for Figure 4.

We also extracted fluxes for each knot (and some sub-knot regions) in the jet. 
These points are listed in Table 2, and plotted along with the best-fit power
laws in Figure 5.  As can be seen, the single power law model is a good fit for
most regions, although a few show signs of gradual curvature in the optical. 
Table 2 also defines the regions used for each knot.  We also applied an
aperture correction of 1.107 (\S 2.2) to both the radio and optical data.

\subsection{Overall Spectral Morphology}

In the optical, all knots in the jet are well fit by single power laws. 
Evidence for spectral curvature in the optical can, however, be seen in D-E,
D-M, A shock, B-1, B-2 and C (Fig. 5).  In all of these cases the curvature
appears to be very gradual and in the sense of spectral steepening towards the
blue, as expected in all standard models of synchrotron emission (see \S 4).  

The jet average is $\langle \alpha_o \rangle \approx 0.9$, slightly flatter
than reported in MRS96 ($\langle \alpha_o \rangle \approx 1$), but
significantly flatter than reported in BSH91 and SBM96 ($\langle \alpha_o 
\rangle \sim 1.3$).  We will examine possible reasons for this in \S 5. We
derive $\langle \alpha_{ro} \rangle \approx 0.67$, very similar to values
observed by previous workers (BSH91, MRS96, SBM96).

Bright regions of the knots have somewhat flatter optical spectra than other
regions of the jet, with $\alpha_o \sim 0.5 - 0.8$ near flux maxima, and
substantial steepening to $\alpha_o \sim 0.8 - 1$ immediately downstream.  Two
regions are particularly flat ($\alpha_o \sim 0.5 - 0.6$): the flux maxima of
knots HST-1 and D-East.   Regions near the flux maxima in E, I and A have only
slightly steeper spectra ($\alpha_o \sim 0.6-0.7$).  By comparison, the
inter-knot regions show significantly steeper optical spectra ($\alpha_o \sim
1.0 - 1.2$), although our signal to noise in those regions is somewhat low and
our errors are correspondingly larger.

 Much smaller variations are observed in $\alpha_{ro}$ than $\alpha_o$. 
Slightly steeper values ($\alpha_{ro} \approx 0.70$) are seen in most regions
of the inner jet (interior to knot A), with slightly flatter values
($\alpha_{ro} \approx 0.66$) observed near the flux maxima of knots HST-1,
D-East, E and F.  By contrast, in the outer jet, a very different pattern is
observed: $\alpha_{ro}$ is remarkably stable, varying by at most $\pm$ 0.03,
with no apparent correlation between jet flux and $\alpha_{ro}$.

In most regions of the jet, the optical spectrum is somewhat steeper than the
radio-optical continuum.  This can be seen on Figures 3c and 4c, where runs of
$\alpha_o - \alpha_{ro}$ are plotted. Values of $\alpha_o - \alpha_{ro}$
average $\sim 0.1$ in bright regions in the inner jet, and $\sim 0.2$ in the
outer jet.   The flux maximum regions of HST-1, D-East, E, I and A, show no
evidence of radio-optical curvature (in fact at their flux maxima HST-1 and D
have $\alpha_o$ somewhat flatter than $\alpha_{ro}$, but at only the $2\sigma$
level).  Significantly larger values of $\alpha_{ro}-\alpha_o$ are observed
downstream of the flux maxima in each knot, indicating increased curvature. 
Interestingly, we observe a nearly constant value of $\alpha_o-\alpha_{ro}$ in
knot regions outside of flux maxima, with no trend for increasing curvature as
the distance from a local maximum increases.  

Perhaps the most glaring feature seen in Figures 2 and 4 is the oscillation of
spectral indices, and the overall anti-correlation of $\alpha_o$ and
$\alpha_{ro}$, observed in the outer jet. As can be seen, the $\alpha_o$ and
$F_I$ maps track each other quite  well but appear almost anticorrelated with
$\alpha_{ro}$, in the sense that the flattest $\alpha_o$ points also represent
the brightest points, but those same points have the steepest values of
$\alpha_{ro}$.

The above discussion indicates that the spectral morphology of the M87 jet is
highly complex.  It is worthwhile to discuss the jet's spectral morphology and
the relationship of the optical and radio spectral indices in greater detail,
isolating each jet region in turn.   The implications of various features of
the jet's spectral morphology will be discussed in \S 6.

\subsection{Spectral Morphology: Inner Jet}

We turn first to the inner jet, which has a knotty morphology, dominated by
five discrete regions.  Proceeding outwards, the first knot is HST-1, $1''$
from the nucleus.  At $0.15''$ resolution, HST-1  consists of a bright peak
with fainter regions downstream.  At higher resolution it breaks up into
several components, which appear to move at speeds up to $6c$ (Biretta et al.
1999, BSM99).   Near the optical flux peak are local minima in both
$\alpha_o$ and $\alpha_{ro}$, but the beginning of the flat $\alpha_o$ region
appears to be $0.15''$ upstream from the beginning of the flat $\alpha_{ro}$
region.  Given the proximity to the central nuclear source and the location of
a nearby dust lane,  this result should be confirmed in future studies.

The brightest knot in the inner jet is knot D, located $2.5-4''$ from the
nucleus.  Knot D can be separated into three regions: the flux maximum, called
D-East, the southern `twist', called D-Middle, and the western edge, called
D-West.  Within this knot are superluminal components moving at speeds ranging
from $2-5 c$ seen in both the optical (Biretta et al. 1999a, BSM99) and radio
(Biretta et al. 1995). The $\alpha_o$ map tracks closely the distribution of
optical flux all throughout knot D, with the flattest spectral indices observed
in a compact ($\sim 0.4''$) region at the flux maximum.  Less prominent local
minima, are seen near the flux maxima in D-Middle and D-West.  The distribution
of $\alpha_{ro}$ is somewhat different:  the minimum in $\alpha_{ro}$ occurs
$0.15''$ upstream of  the extrema in $\alpha_o$ and $F_I$; the increase in
$\alpha_o$ which follows the flux maximum also appears to lag the increase in
$\alpha_{ro}$ by this same amount.  These two features show up most
dramatically in the run of $\alpha_o - \alpha_{ro}$ (Fig. 3c).  The
$\alpha_{ro}$ image does not show significant local extrema near the locations
of either D-Middle or D-West, but some slight flattening trend is seen over
this region.  Regions closer to the jet axis appear to have flatter values of
$\alpha_o$ than the northern or southern edges, but this is not at all clear
for $\alpha_{ro}$.

At about $6''$ from the nucleus we observe knot E, which is barely resolved
along the jet axis at $0.15''$ resolution (at higher resolution two
condensations are seen; SBM96).  Speeds of $\sim 3.5 c$ have been detected in
our HST monitoring program (BSM99).  Knot E does not exhibit the sharp spectral
changes seen in knot D, but nevertheless a flatter-$\alpha_o$ region is seen
centered roughly on the flux maximum. There is also a local flattening in
$\alpha_{ro}$ which appears more extended along the jet axis than the
flat-$\alpha_o$ region; however $\alpha_{ro}$ reaches minimum $0.25''$ upstream
of $\alpha_o$. Unlike in knot D, however, the downstream ends of the flat
$\alpha_o$ and $\alpha_{ro}$ regions appear spatially coincident.

Knot F, located $8-9''$ from the nucleus, appears rather amorphous at $0.15''$
resolution but breaks up into two diffuse blobs at higher resolution.  Radio
monitoring has detected speeds near $c$ in this region (Biretta et al.  1995),
but our optical monitoring has not yet detected proper motion (BSM99).  Knot
F's spectral morphology appears rather subtle (Figures 2a-b, 3).  Both the
$\alpha_o$ and $\alpha_{ro}$ images show flatter spectral indices closer to the
jet axis than along the northern or southern axes. Neither spectral index map
shows strong local extrema, but instead both show a flat spectrum region
roughly coincident with the rather broad flux maximum region.  At its upstream
end, $\alpha_{ro}$ appears to lag $\alpha_o$ by $0.17''$, but the situation is
reversed at the downstream end, where the flat $\alpha_o$ region terminates
$0.2''$ before the termination of the flat $\alpha_{ro}$ region.  

Knot I, located $11''$ from the nucleus, appears to have motions near $c$ from
HST monitoring (BSM99).  Examination of Figures 2a and 2e reveals a fairly
compact flat-$\alpha_o$ region centered on the flux maximum.  However, the
changes in $\alpha_{ro}$ are not well localized to the flux maximum region.

\subsection{Spectral Morphology:  Outer Jet}

As has been noted by many authors, the morphology of the outer jet is quite
different from that of the inner jet.  The outer jet features four knot
complexes, traditionally called A ($12-14''$ from the nucleus), B ($14-16''$
from the nucleus; {\it n.b.} neither the downstream end of knot A nor the
upstream end of knot B has a sharp ``edge''), C ($17-19''$ from the nucleus)
and G ($>19''$ from the nucleus). 

Looking at the outer jet in Figures 2a-b and 4, one is immediately struck by
the strong anti-correlation between $\alpha_o$ and $\alpha_{ro}$.   This
anti-correlation begins in the inter-knot region between I and A, $11.4-11.8''$
from the nucleus, where we observe a flattening in $\alpha_{ro}$ with a
possibly double structure, corresponding to a region with steep $\alpha_o$.
Further downstream, the flux maxima in A, B and C all correspond to regions 
with flat $\alpha_o$ but steep $\alpha_{ro}$ values, as do the `filamentary'
structures in this region (although the departures are smaller than for the 
flux maxima).  Also agreeing with this trend are the steeper than average
values of $\alpha_o$ and flatter than average values of $\alpha_{ro}$ seen in
the B-C interknot region.  A close look at Figures 2a-b and 4, in fact,
reveals, that local minima in $\alpha_o$ coincide very closely with local
maxima in $\alpha_{ro}$ and vice versa; moreover the locations of increases in
$\alpha_o$ also coincide very closely with decreases in $\alpha_{ro}$ and vice
versa.

\section{Synchrotron Spectrum Models}

We fitted synchrotron spectrum models to the integrated radio through X-ray
fluxes for knots A, B and D as well as the larger A+B+C complex, and on a pixel
to pixel basis to radio through optical data for the entire jet, using programs
written by C. Carilli and J. P. Leahy (Carilli et al. 1991, Leahy 1991).  In
these fits we include X-ray data in our analysis for comparison with the total
fluxes of knots  A, B and D, where  X-ray emissions have been detected
(Schreier et al. 1982, BSH91).  For knot A, we use X-ray fluxes from a January
1998 ROSAT HRI observation (Harris et al. 1998).  For knots B and D, fluxes
have not yet been extracted from ROSAT data, so we used 1 keV fluxes from {\it
Einstein} (BSH91){\footnote{Recent observations of the M87 jet by {\it Chandra}
have in fact detected X-ray emission from all knots in the jet (Marshall et al.
2001).}}.   The conversion of X-ray counts to flux for
all regions was done assuming $\alpha_x=1.3$ and log $N(H) = 20.38$ (BSH91). 
Even though the X-ray data for B and D are not contemporaneous, this will have
only a small effect on $\alpha_{ox}$. A 50\% change in flux (3$\times$ as
large as the variations in A documented by Harris et al. 1997, 1998) would only
produce a change of $\Delta \alpha_{ox} = 0.10$.

The models we fit were:

(1) A Jaffe \& Perola (1973) model (hereafter JP), which allows for continuous
isotropization of the pitch-angle distribution of the electron population with
time, but no further particle injection.

(2) A Kardashev-Pacholczyk model (Kardashev 1962, Pacholczyk 1970; hereafter
KP), which does not allow pitch-angle scattering, and therefore allows the
development of a high-energy ``tail'' of particles at oblique pitch angles.

(3) A ``continuous injection'' (hereafter CI) model (Heavens \& Meisenheimer
1987, Meisenheimer et al. 1989) under which a power-law distribution of
relativistic electrons is continuously injected.  This model was originally
developed for the case of an unresolved source. Meisenheimer et al. (1989) and
Heavens \& Meisenheimer (1987) have shown that this model applies when there
exists within the telescope beam an unresolved zone of continuous high-energy
particle injection ({\it e.g.}, a shock front).

Each model can be characterized by an injection index $\alpha_{in}$ and a break
frequency $\nu_b$.   The differences between the JP, KP and CI spectrum show up
most strongly blueward of the break.  The JP spectrum has the fastest cutoff,
the CI spectrum has the least steep break, and the KP model's cutoff is
intermediate between the two.  Importantly, the overall trends fit by each
model are identical ({\it e.g.,} if a region has high $\nu_b$, this will be
true under all three models).  Therefore, the pixel-by-pixel fits were done
assuming only the KP model. 

In fitting integrated fluxes, we tried only $\alpha_{in}= \alpha_{ro}$, as well
as $\alpha_{in}=\langle \alpha_{in} \rangle$  (with the mean derived from the
$\alpha_{in}$ map).  We did not attempt a $\chi^2$ minimization for these fits
because all of those models have relatively fixed general characteristics;
moreover, as we will emphasize, because of the large gaps which exist in the
mid to far IR as well as UV, there are significant limits to what this
procedure can do.  The results of these fits are shown in  Figures 6a-e, and
the derived $\nu_b$, $\alpha_{in}$ and  $\chi^2$ values are summarized in Table
3.

Inspection of Figure 6 reveals a striking result: independent of whether the
X-ray emissions are synchrotron in nature, the CI model is decisively ruled out
by our data because it overpredicts by large factors (5-50) the X-ray fluxes. 
The CI spectrum could be made to work for knot A if the X-ray emitting
electrons occupy a region about 1/5 the volume of knot A in the optical, but
greater reductions are necessary for knots D (factor 8) and B (factor 50). 
This topic will be discussed further in \S 6.

By comparison, both the JP and KP models fit the data reasonably well; our data
cannot discriminate between these models because there are no data in the break
region (the EUV).  An interesting point in favor of a synchrotron nature for
the X-ray emissions is that in knot D, JP and KP models fit to only the radio
through optical data, predict the X-ray emissions to within a factor $\sim 2$.

Pixel by pixel fits were done only to the radio through optical data. In
Figures 2c and 2d, we show (respectively) maps of $\nu_b$ and $\alpha_{in}$ for
the KP model only.  Note that good quality fits were only achieved in knot
regions and so we have clipped these images so that they do not include
interknot regions.  For Figure 2c, the color scale runs from $10^{14.5}$ to
$10^{16}$ Hz, while the color scale for Figure 2d is identical to that used for
Figure 2b.

We derive break frequencies $\sim 10^{16}$ Hz in most inner jet regions.  Note
that values $\sim 10^{16}$ Hz should be treated as guidelines and/or lower
limits.  Due to the gradual curvature of synchrotron spectra ({\it e.g.},
Figure 6), a lack of significant radio-optical curvature makes fitting values
of $\alpha_{in}$ and $\nu_b$ difficult, and in such cases the code used outputs
$\nu_b$ values 10 times higher than the highest frequency for which data are
present (here $10^{15}$ Hz).  Somewhat lower break frequencies ($3-10 \times
10^{15}$ Hz) were found in knots A, B and C, with higher break frequencies in
flat-spectrum regions near knot maxima.  Note that we do not quote break
frequencies in inter-knot regions because the fits in those regions were poor
due to their low signal to noise.  These values agree well with those we find
for the integrated fluxes of knots D, A and B, but with the addition of X-ray
data.

The $\nu_b$ (Figure 2c) and $\alpha_o$ (Figure 2a) maps are extremely
well correlated, with high break frequencies in the flattest spectrum
regions, and low break frequencies in the fainter, steep spectrum
regions.  Also, the $\alpha_{in}$ map (Figure 2d) largely correlates
with the $\alpha_{ro}$ map (Figure 2b).  Both of these are to be
expected given the rather large gap (4 decades in frequency) between
the radio and optical points.  Note that also from a physical point of
view one expects $\alpha_{ro}$ and $\alpha_{in}$ to be correlated
unless there is significant low-frequency radio structure which is
self-absorbed by 15 GHz.  We do not have the data to test for this
possibility but it is somewhat beyond the scope of this paper.

\section {Comparison With Previous Work}

We report average values of $\alpha_{ro}$ which are comparable to those
reported by other authors.  The average value of $\alpha_o$ we find (0.9) is,
however, flatter than values reported by previous authors, which range from
$\langle \alpha_o \rangle \approx 1.0$ (MRS96) to $\langle \alpha_o \rangle
\approx 1.3$ (BSH91 and SBM96). It is worthwhile to look at the likely reasons
behind these discrepancies.  Since the regions used to  obtain fluxes (as well
as the calibration process) varied in each case, the comparison is somewhat
difficult.  We therefore will speak  in rather broad terms about the global
picture, and concentrate on the A+B+C region, for which all previous workers
except for SBM96 have compiled fluxes, summarized in MRS96.

As already mentioned, Figures 6d and 6e contain our fluxes for the knot A+B+C
region as well as the synchrotron spectrum fits for this region.  We have
overplotted on Figures 6d-e the other fluxes quoted by MRS96; particularly
relevant to this discussion is Figure 6e which zooms in on the optical.  As can
be seen, our flux values are well in accord with all but two teams' fluxes. 
The discrepant values are from BSH91 and Killeen et al. (1984).  We cannot
throw any additional light on why those values are discrepant; we refer the
reader to the discussion in MRS96 on this point.  Moreover, the overall value
of spectral index one derives from our values is easily seen to be identical
within the errors to what one would derive from all previous points except the
IUE values (Perola \& Tarenghi 1980), which may be discrepant because of poor
galaxy subtraction.  Those values bear some mention as they account for most of
the curvature in the models fit by MRS96 to the data for A+B+C, and also have
by far the largest error bars.  Also, those points, plus those of BSH91, are
the only ones consistent with a steep value for $\alpha_o$ in the A+B+C region.

As can be seen by comparing our Figures 3a and 4a with Figure 3 of MRS96, our
values and those of MRS96 for $\alpha_o$ are completely consistent if a
constant offset of $\Delta \alpha_o \approx 0.1$ is applied, and fine structure
on scales smaller than an arcsecond is ignored.  It is likely that the main
reason for the constant offset is contamination from the galaxy, which would
tend to redden the spectrum, since the mean color of
starlight from M87 is significantly redder than the jet ($R-I$ color of the
galaxy's light averages 0.78 mag, Boroson \& Thompson 1991; as compared to $r-i
=$ 0.16 mag for knot A of the jet, BSH91).  This is perhaps not entirely
unexpected given the much higher resolution of HST compared to ground-based
telescopes.  It is hard to imagine other significant color-based or systematic
effects given that the overall flux scales agree quite well (Figure 6e).  A
second factor in this respect, however, could be the larger number of bands in
our data set (7 compared to 3) which  allows us to better fit spectral indices
as well as see curvature.

We fit models which are somewhat different than those fit by MRS96. In
particular we do not make assumptions about the nature of a spectral cutoff at
high frequencies.  Therefore it is somewhat confusing to compare the values of
{\it break} frequency we derive to the values of {\it cutoff} frequency they
derive.  It is also important to note that the comment made in MRS96 that there
is evidence for a steeper cutoff than predicted by Fermi acceleration was based
only on the IUE points and so this needs to be verified (we have since
obtained shorter wavelength data with STIS; those data will be the subject of a
later paper).  Given all of this plus the somewhat flatter values of $\alpha_o$
we find, it is therefore not surprising that our values for $\nu_b$ are higher
than their $\nu_{cutoff}$ values.  In comparing the two tracks however (cf.
Figure 8b in MRS96 and our Figure 2c) it is important to note that,
particularly in the outer jet, the overall trends are exactly the same except
for the resolution difference.

The values reported by SBM96 were based on pre-COSTAR HST data.  A significant
number of those datasets are low S/N and the spectral index fits in SBM96 were
not weighted by signal to noise. Moreover, the regions used by SBM96 contain
large non-jet portions because they were rectangular regions defined on an
image where the jet was inclined by roughly 21$^\circ$ to the x axis.  The data
used here are considerably higher signal to noise in every band, and the jet is
on the x axis; in addition, we do not have to deal with the residuals caused by
deconvolving the pre-COSTAR PSF in each band.  We have refit spectral indices
to the quoted flux values for various jet regions, using only the F140W, F220W
and F372M datasets (the only ones with reasonably good S/N  over large parts of
the jet). When this is done we obtain optical spectral indices considerably
flatter than those obtained by SBM96 (probably due to our exclusion of the
lower signal to noise data, which also would have had poorer galaxy
subtraction), and indeed in bright regions such as knot A we obtain results
comparable to within the errors.  We have also resampled those images to
$0.0758667''$/pix, smoothed them to $0.15''$ resolution, and fit a spectral
index image using the same program described in \S 3.  The result is extremely
noisy (with error bars much larger than in our data), but spectral indices
considerably flatter than 1 are found in bright regions; in fact near the flux
maxima of A and D-East spectral indices $\sim 0.4 - 0.7$ are found, with
internal error bars $\sim \pm 0.2$.  This tends to support the reliability of
our results.

Finally, there is also a large discrepancy between our spectral index values
and those quoted by BSH91.  As we have already commented (and as previously
noted by MRS96), however, the flux scale used in BSH91 appears to be rather at
variance with that of other workers.  We do not have significant new light to
throw on the flux scale issue but given the overall good agreement of our flux
scale with that of  previous workers we are suspicious of the spectral index
values  quoted by BSH91.

\section {Discussion}

We have presented data which give the first high-resolution picture of the
near-infrared to ultraviolet spectrum of the M87 jet, and indeed represent the
first dataset of this type for any extragalactic jet. We have given a detailed
discussion of the spectral morphology of the jet and the $\alpha_o-\alpha_{ro}$
relationship in \S\S 3.3 and 3.4, and discussed synchrotron spectrum model fits
in \S 4.  These allow us to place significant constraints on the jet structure,
the nature of the X-ray emission, and also the extent, need and nature of
particle acceleration in the M87 jet.  We discuss each of these topics in turn.

\subsection {Nature of the X-ray Emission}

Our data are fully consistent with a synchrotron origin for the X-ray emission
in the jet (as proposed by BSH91).  The observation of variability in the X-ray
emission from knot A and the nucleus (Harris et al.  1997, 1998), rules out a
thermal bremsstrahlung nature for the X-ray emission.  But neither our data nor
the observed variability are completely sufficient to rule out an
inverse-Compton nature for the X-ray emission.  

If one assumes that the X-ray, optical and radio synchrotron emitting electrons
in knots A, B and D are co-spatial, the CI model of Meisenheimer et al. (1989)
is ruled out because it overpredicts by large factors the observed X-ray
emission.  Thus particle injection or acceleration can occur only in discrete
sites occupying a small fraction of the jet volume. This agrees with the
compactness required by the detection of X-ray variability in knot A on
timescales of months (Harris et al. 1997, 1998).  The CI model can be made to
work for knot A if the X-ray emitting electrons in that knot occupy a region
about 1/5 the projected area of knot A in the optical.  Similar arguments can
be made for knots D (factor 8) and B (factor 50), although their X-ray
detections are less secure. 

Measurement of the X-ray spectral index $\alpha_x$ would allow us to
discriminate between the synchrotron and inverse-Compton models. 
Inverse-Compton X-ray emission would have a much flatter X-ray slope ($\alpha_x
\sim 0.7$, comparable to the jet's $\alpha_{ro}$ and similar to the average
hard X-ray slopes of low-energy peaked BL Lacs (LBL) found by ASCA, see Kubo et
al. 1997), than both the JP and KP synchrotron models, which predict $\alpha_x
\sim 1.5-2$ and $\alpha_{ox} < \alpha_x$, as seen  in the X-ray spectra of most
high-energy peaked BL Lacs (HBL; Perlman et al.  1996, Sambruna et al. 1996).  

Unfortunately, because of the multi-component nature of the X-ray emission from
the M87 region (including cluster, galaxy and jet components), current X-ray
data do not allow us to place good limits on the shape of the X-ray slope of M87's
jet.  This could not be done from the {\it Einstein} HRI or ROSAT HRI datasets
due to their poor (at most two-channel) energy resolution.  Higher energy
telescopes (ASCA, SAX and RXTE) have tried to determine the relative
jet+nuclear X-ray emissions but with at best variable success due to their much
poorer ($\sim 1-3'$) angular resolution.  Matsumoto et al. (1996) and Allen et
al. (2000) both find some evidence of a hard X-ray tail in ASCA data but with
large error bars on the spectral shape ($\alpha = 0.4^{+0.4}_{-0.5}$ between
2-10 keV).  Similar constraints ($\alpha_x=0 \pm 1$), have recently been
obtained from SAX data (Guainazzi \& Molendi 1999). However, a negative
detection was obtained by RXTE (Reynolds et al. 1999), a non-imaging detector,
which places a lower limit of $\alpha_x = 0.7$ on the spectral index of knot A,
and $\alpha_x=0.9$ on the core.  Comparison of the fluxes obtained in the two
detections plus the upper limits, however, does reveal evidence of variability
(Guainazzi \& Molendi 1999, Reynolds et al. 1999).  Thus current X-ray data
cannot constrain the spectral shape of any component in the M87 jet, but they
do constrain the nature of the emission to nonthermal (synchrotron or
inverse-Compton) mechanisms because of the observed variability.

\subsection {Jet Structure}

As shown in Figures 2-4 and \S 3, the relationship between $\alpha_o$ and
$\alpha_{ro}$ in the M87 jet is rather complex.  In the inner jet, the two
appear correlated in most regions if a lag is applied to one or the other,
while they are essentially anti-correlated in the outer jet.  These trends are
further evidence that it is impossible to model the M87 jet as an axially
homogeneous jet where radio and optical synchrotron emitting electrons are
completely co-spatial.  Instead, they support a model for the structure of the
jet presented by P99, whereby the energy spectrum of electrons in the jet, and
magnetic field direction, vary significantly with distance from the jet axis.
In the P99 model, higher-energy electrons are concentrated closer to the jet
axis, where the magnetic field is most strongly affected by shocks and other
disturbances, while lower-energy electrons occupy mainly the jet ``sheath'' at
greater distances from the jet axis.  At any one point, the observed spectral
indices include both core and sheath contributions, with the core dominating in
the optical and the sheath dominating in the radio.  This model predicts that
$\alpha_o$ and $\alpha_{ro}$ should not vary together.  It also predicts that
some regions might have flatter {\it observed} values of $\alpha_o$ than
$\alpha_{ro}$, since different regions can predominate in different bands.  This
is in fact observed at the flux maxima of HST-1 and D.

The poor correlation between $\alpha_o$ and $\alpha_{ro}$ in the outer jet is
somewhat more surprising given the large degree of similarity between the radio
and optical polarized structure in this region (P99).  However, in the absence of
strong shocks it is still entirely possible for energy stratification to occur
without observing significant differences in polarimetry.  Indeed, we should point
out that the mixing of previously stratified layers would not be expected to occur
in a short, linear shock, and would likely produce turbulence which would be
visible in the form of eddies (not observed). Moreover,  the optical polarized
structure of the outer jet, while much more similar to that seen in the radio,
still does show evidence that shocks begin in the jet interior, particularly in
the pre-shock regions of A and C (refer to Figures 2, 5 and 6 of P99).  The
details of the spectral index maps we find indicate that regions of the jet closer
to the axis probably  have somewhat steeper values of $\alpha_{ro}$ and hence
$\alpha_{in}$, than the sheath.

The finding that the X-ray emissions must fill a volume significantly smaller
than the lower-energy emission also agrees with the stratification proposed by
P99.  It should be noted that because of the observed variability, this result
is independent of whether the X-ray emission mechanism is synchrotron or
inverse-Compton.

\subsection {Is {\it In Situ} Particle Acceleration Necessary?}

The issue of whether {\it in situ} particle acceleration is required for the
M87 jet, is complex and not well explored.  As explained by BSH91, particle
acceleration is required if the magnetic field is near the equipartition value
of $B\sim 200 \mu$G.  This is underlined by our data,  which yield
``synchrotron ages'' for knots D, A and B of 60-100 years under the JP model,
and somewhat higher (110-280 years) under the KP model, similar to values
calculated by BSH91 under different assumptions.  However, the radiative
lifetimes of synchrotron emitting particles vary strongly with magnetic field
($\propto B^{-3/2}$), so that if the magnetic field is significantly weaker
than the equipartition value (Heinz \& Begelman 1997), it is not unfeasible to
reconcile the radiative lifetimes of synchrotron emitting electrons with the
length of the jet ($\sim 2$ kpc).

The local variations in optical spectral index and break frequency we find, as
well as the optical, radio and X-ray variability observed in HST-1, D and A
(Biretta et al. 1999, Zhou 1998, Harris et al. 1997, 1998), represent strong
evidence for particle acceleration near the flux maxima of those regions.  Such
particle acceleration is in fact expected if knots D and HST-1, where several 
variable superluminal components have been found in both HST and VLA monitoring,
are analogous to flaring regions in blazar jets.  Moreover, the details of the
optical polarimetry in those and other regions (P99), where we observe magnetic
field vectors to be strongly compressed transverse to the jet direction, creates
all the conditions necessary for particle acceleration.  

If indeed the observed changes in spectral index are a product of particle
acceleration, then Figures 3-4 represent a time history of particle
acceleration and synchrotron aging in the jet in regions where  the implied lag
$<$ light-crossing time of the jet.  This interpretation allows us to comment
on the nature of particle acceleration in the M87 jet, as well as the
acceleration and cooling timescales, because it transforms the observed
differences in the runs of $\alpha_o$ and $\alpha_{ro}$ versus distance into
time lag information.  In two knot regions it appears that variations in
$\alpha_o$ lag variations in $\alpha_{ro}$.  In D-East, a lag of $0.15''$ is
observed, which translates to about 7.5 years using the observed speeds
$\sim 5c$ in this region (BSM99).  The somewhat larger lag of $0.25''$ seen in
knot E translates to about 14 years given the speeds in BSM99.  There are also
two knot regions (HST-1 and F) where we observe the opposite trend: $\alpha_o$
leading $\alpha_{ro}$.  In the case of knot HST-1 the observed lag translates
into a period of $\sim 7$ years given the speeds in BSM99 (greater precision is
not possible given the nearness of the point source), while for knot F we
cannot translate the lags into timescales directly since the velocity fields
are not well known in those knots.

If particle acceleration is present in the M87 jet, the mechanism is not well
constrained.  Several mechanisms have been proposed, including magnetic
reconnection (Lesch \& Birk 1999) and Fermi acceleration in knots
(Meisenheimer, Yates \& R\"oser 1997). The structure of the jet, which contains
many shock-like, flat-spectrum knots (see also P99), argues against a major
role for magnetic reconnection.  In contrast, the observation of sharp changes
in optical magnetic field position angle in these same flat-spectrum knots
(P99), suggests compression of the magnetic field in the jet interior at these
positions, more consistent with shocks.

The observed spectral variations near knot maxima in the inner jet are consistent
with models for flares produced by shock-induced particle acceleration. As Kirk,
Rieger \& Mastichiadis (1998) have shown, the evolution of spectral index in a
given band during a flare depends on the relationship of the acceleration and
cooling timescales for electrons which emit in that band.  Where the acceleration
timescale is much shorter than the cooling timescale, higher-energy emissions
should lead lower-energy emissions (Georganopolous \& Marscher 1998, Takahashi et
al. 1999), so that a flare will propagate from the highest energies to lower
energies.  But if the cooling and acceleration timescales are close to equal near
where the peak in $\nu F_\nu$ is, the relationship between spectral changes is
complex.  In that case, at low energies, where the acceleration timescale is much
faster than the cooling timescale (note that $t_{cool} \propto E_\gamma^{-1/2}$
for synchrotron radiation), their model predicts a spectral flattening in advance
of the flux maximum, peaking slightly before flux maximum, followed by a decrease
to values somewhat below the jet's nominal value after maximum flux, and then a
return to the nominal value after the disturbance has passed.  But at high
energies, where $t_{cool} \sim t_{acc}$, a spectral {\it steepening} is predicted
in advance of the flux maximum, followed by a hardening as the number of newly
accelerated electrons build up.  Under this model (assuming no injection of fresh
particles), flares represent the reacceleration of old particles within the jet,
so they will propagate from low to high energies, {\it i.e.} observed spectral
changes at higher energies will lag those at lower energies.  We appear to be
observing {\it both} situations at different locations in the optical emission
from the M87 jet, perhaps not dissimilar from what has been seen (on shorter
timescales) in blazars; in particular recent observations of PKS 2155$-$304
several examples of both types of flares were seen within a week of observation
(Sambruna et al. 1999).

\subsection{Comparison to Radio-Optical Spectral Analysis for Other Jets}

With the above data and analysis in mind we can compare our findings to those
of other teams on various optical jets.  Unfortunately high-quality data
(either ground-based or HST) are few and far between, and only exist on a few
sources.

The only other radio/optical jet to have been imaged polarimetrically with HST is
3C 273.  Comparing the HST and radio polarimetry for its jet (Thomson, Mackay \&
Wright 1993, Conway et al. 1993), 3C 273 seems to show more similarity between
radio and optical polarized structure than does the M87 jet (P99), despite having
a large number of knots.  However, a deep comparison of its  radio and optical
polarized structure is currently impossible because of the  low signal to noise of
the existing, pre-COSTAR FOC polarimetry data. Further, a deep analysis of its
optical spectrum has not yet been done, so we need to rely on ground-based data to
compare the polarized and spectral index maps.  Neumann, Meisenheimer \& R\"oser
(1997) and R\"oser \& Meisenheimer (1991) have carried out a deep analysis of the
near-IR to optical spectrum of the 3C273 jet (note that the latter paper also
contains ground-based polarimetry, which is also useful), which fortunately is
largely unaffected by galaxy subtraction issues (unlike ground-based work for
M87).  They find considerably more variation in $\alpha_o$ than $\alpha_{ro}$ for
3C 273, as we do for M87, and also find that 3C 273's jet is significantly
narrower in the optical than in the radio (as we do for M87; see SBM96).  A more
recent, preliminary comparison of the deep HST/WFPC2 and radio images (R\"oser et
al. 1997, Jester et al. in prep.) show some other differences between the jet's
optical and radio structure: the optical jet appears to stop $1''$ short of the
terminal radio hotspot, and the optical and radio flux maxima of knot B are not
spatially coincident. Interestingly, they also find somewhat of an
anti-correlation between optical polarization and spectral index structure, with
low polarization and fairly flat spectral indices $12-14''$ from 3C 273's nucleus,
but higher polarization and much steeper spectral indices further out.  More
recently, {\it Chandra} has imaged the 3C273 jet, finding a rather different
morphology in X-rays than either the optical or radio: one knot's flux maximum
appears displaced, and the optical emission declines drastically in the knot D-H
region, where the strongest radio emission is seen, analysis of the knots' spectra
reveal inconsistencies between the spectra of downstream knots and the predictions
of SSC models (Marshall et al. 2000).  Further analysis and deeper comparisons of
the X-ray, optical and radio structure of the 3C 273 jet will no doubt give us
considerable insight into its physics.

Lara et al. (1999) find a steep optical spectrum for 3C264's jet, and a
broadband shape consistent with either the CI model or a significantly
sub-equipartition magnetic field (Heinz \& Begelman 1997), and that it is
unlikely that the observed optical jet emissions are inverse-Compton in
origin.  They invoke the lack of well-defined knots in the 3C264 jet as
evidence against Fermi acceleration in shocks. This, plus the much earlier
deceleration found for the 3C264 jet (Lara et al. 1999, Baum et al. 1997), are
key distinctions between it and M87, which may make 3C264 more amenable to the
CI scenario than M87, where several bright shocks are observed.  High signal to
noise X-ray or UV observations, as well as deeper HST imaging data in several
bands (the HST data analyzed by Lara et al. were 300s snapshots), as well as
HST polarimetry, would go a long way towards resolving the mechanism which
dominates in the 3C264 jet.

Schwartz et al. (2000) and Chartas et al. (2000) have discussed the discovery
with {\it Chandra} of an X-ray jet in the quasar PKS $0637-752$, and the
subsequent identification of optical emission from three knots in its jet in
deep HST images of the field.  This object is at much larger distances than 3C
273, let alone M87 or 3C 264, making detailed comparisons of structure
difficult.  However, comparisons of the structure in the radio and optical
reveal large-scale correspondence but tantalizing hints of differences on
scales of $\sim 0.2''$ which require deeper investigation.  The broadband
spectra of the knots are also inconsistent with the predictions of simple SSC
models, although Comptonized emission from CMB photons remains a possibility
(Tavecchio et al. 2000)

One final object for which such analysis can in future be done is 3C 66B.  Sparks
et al. (in prep.) are analyzing radio-optical spectral data for this object, which
also shows interesting radio-optical spectral differences.

By discussing the kinetic energy budget of optical jets, Scarpa \& Urry (1999)
find that their observed properties require either fairly high bulk Doppler
factors $\Gamma \sim 5$ and viewing angles $\lsim 25^\circ$ (if equipartition is
maintained), or magnetic fields significantly less strong than equipartition
(whereupon lower values of $\Gamma$ or larger viewing angles are possible).  This
is in agreement with the findings of Heinz \& Begelman (1997) for M87, and
consistent with our results.  A similar result has been found through comparison
of the properties of dust disks in the 3CR sample (Sparks et al. 2000).

\section{Scope for Future Observations}

The above sections underline the importance of future observations. Since our data
indicate break frequencies significantly in excess of $10^{15}$ Hz,
high-resolution imaging observations in the far UV and X-rays are now required to
localize the peak of the synchrotron emission from bright knots in the jet.  Such
data, if gathered contemporaneously with optical and radio observations, would
allow much stronger constraints on synchrotron emission and particle acceleration
models.  

The {\it Chandra} observations are particularly
critical in this regard. Those observations will yield spatially
resolved (resolution $\sim 0.7''$) spectra in the $0.1-10$ keV region
({\it i.e.,} $2-200 \times 10^{16}$ Hz), allowing direct comparisons of
optical and radio morphologies.  Such comparisons will directly test
the findings of Neumann et al. (1997) that the X-ray emission from
knot A is not coincident with the radio/optical peak, and will allow
us to localize sites of particle acceleration.  The {\it Chandra}
observations will also directly test the nature of the jet's X-ray
emission by measuring the X-ray slope (\S 5.1).

Our data indicate that observations in the far-UV are required
to localize the synchrotron peak of much of the jet.  No imaging 
mission is currently proposed at these wavelengths; however, the 
prospects for detecting M87 in the far-UV are bright, since the $N(H)$ 
along its line of sight is relatively low (although EUVE did not detect
M87; Lampton et al. 1997).

Future monitoring experiments will have unparalleled importance for
models of jet flares in blazars (for a review of the spectral
evolution observed in HBL flares, see Takahashi et al. 1999).  Once
the timescale of X-ray variability is established, the combination of
contemporaneous VLA, HST and {\it Chandra} observations will make it
possible to for the first time map out the broadband spectrum and
polarization of a flaring component, and watch its spectrum decay once
the flare has passed.  Such a study would allow us to fully model the
spectrum and magnetic field configuration in the flaring region.

\begin{acknowledgments}

We wish to thank Eddie Bergeron and Megan Sosey, as well as Shireen Gonzaga,  for
considerable help with the HST data reductions.  We also  wish to thank our
referee, Hermann-Josef R\"oser, for many suggestions which significantly improved
this paper.  JPL thanks the STScI for  hospitality during his sabbatical visit. 
ESP thanks Chris Reynolds,  Sebastian Heinz, Mitch Begelman, Joel Carvalho, Tim
Heckman, Julian Krolik, Eric Agol, Greg Madejski and Jean Eilek for interesting 
discussions.  Research on M87 at STScI is supported by HST grants GO-5941,
GO-7274, GO-7866, GO-8048 and GO-8140.  ESP acknowledges support at Johns Hopkins
University and the University of Maryland, fromfromfromfrom NASA LTSA grant
NAG5-9534/NAG5-9997.

\end{acknowledgments}

\vfill\eject

\centerline {\bf{Figure Captions}}

{\bf Figure 1.} The jet of M87, as seen by HST, in the F300W (0.3 microns,
top), F814W (0.8 microns, middle) and F160W (1.6 microns, bottom) bands.  Flux
levels displayed are from -0.3 to 1 $\mu$Jy/pix for the F300W image, -0.5 to 2
$\mu$Jy/pix for the F814W image, and -1 to 4 $\mu$Jy/pix for the F160W image,
and within these limits the identical color scale (shown at bottom) was used
for all three images. The images have been rotated so that the jet is along the
X axis, and have identical pixel sizes, origins and orientations.  The
traditional names are used for bright knots.

{\bf Figure 2.} Maps of fitted quantities for the M87 jet, with the convolved
F814W image of the jet shown for comparison at the bottom. At top (Figure 2a),
we show the optical spectral index image; second from top (Figure 2b), we show
the radio-optical spectral index image. We also show maps of the synchrotron
break frequency (third from top, Figure 2c) and injection index (fourth from
top, Figure 2d) in the KP model.  Identical color tables were used for all
panels.  The color scale in Figure 2a runs from 1.2 to 0.4, while that in
Figures 2b and 2d run from 0.85 to 0.6, and that for the break frequency map
(Figure 2c) runs from $10^{15}$ to $10^{16}$.

{\bf Figure 3.} Runs of $\alpha_o$ (top, Fig. 3a), $\alpha_{ro}$ (second from
top, Fig. 3b), $\alpha_o-\alpha_{ro}$ (third from top, Fig. 3c) and F814W flux
(bottom, Fig. 3d), plotted as a function of distance along the jet axis.  In
this Figure we show the inner regions of the jet, out to $10''$ from the
nucleus.  These values are {\it flux-weighted}, i.e., flux is summed across the
jet at each pixel in the x direction before spectral indices were computed. 
See \S \S 3 for discussion.

{\bf Figure 4.} Runs of of $\alpha_o$ (top, Fig. 4a), $\alpha_{ro}$ (second
from top, Fig. 4b), $\alpha_o-\alpha_{ro}$ (third from top, Fig. 4c) and F814W
flux (bottom, Fig. 4d), plotted as a function of distance along the jet axis. 
In this Figure we show the outer regions of the jet, $10-20''$ from the
nucleus.  These values are {\it flux-weighted}, i.e., flux is summed across the
jet at each pixel in the x direction before spectral indices were computed. 
See \S \S 3 for discussion.

{\bf Figure 5.}  Fluxes of  knot regions.  The regions are defined in Table 2. 
The error bars include flatfielding, zeropoint and Poisson errors.  Best-fit
power law spectra are shown as lines on these log-log plots, and the spectral
indices are given in Table 2.

{\bf Figure 6.}  Synchrotron spectrum model fits to four regions of the jet:
knot D (Figure  6a), knot A (Figure 6b), knot B (Figure 6c), and knots A+B+C
(Figures. 6d-e).  Along with our data, we plot fits from the JP, KP and CI
models, both including and not including the X-ray point.  In Figures 6d-e, we
show also previous data as tabulated by  MRS96 (see references and discussion
in that paper).   In all plots, data from this work are plotted as diamonds; in
Figs. 6d-e, data taken from MRS96 are plotted as squares. See \S 4 for
discussion.

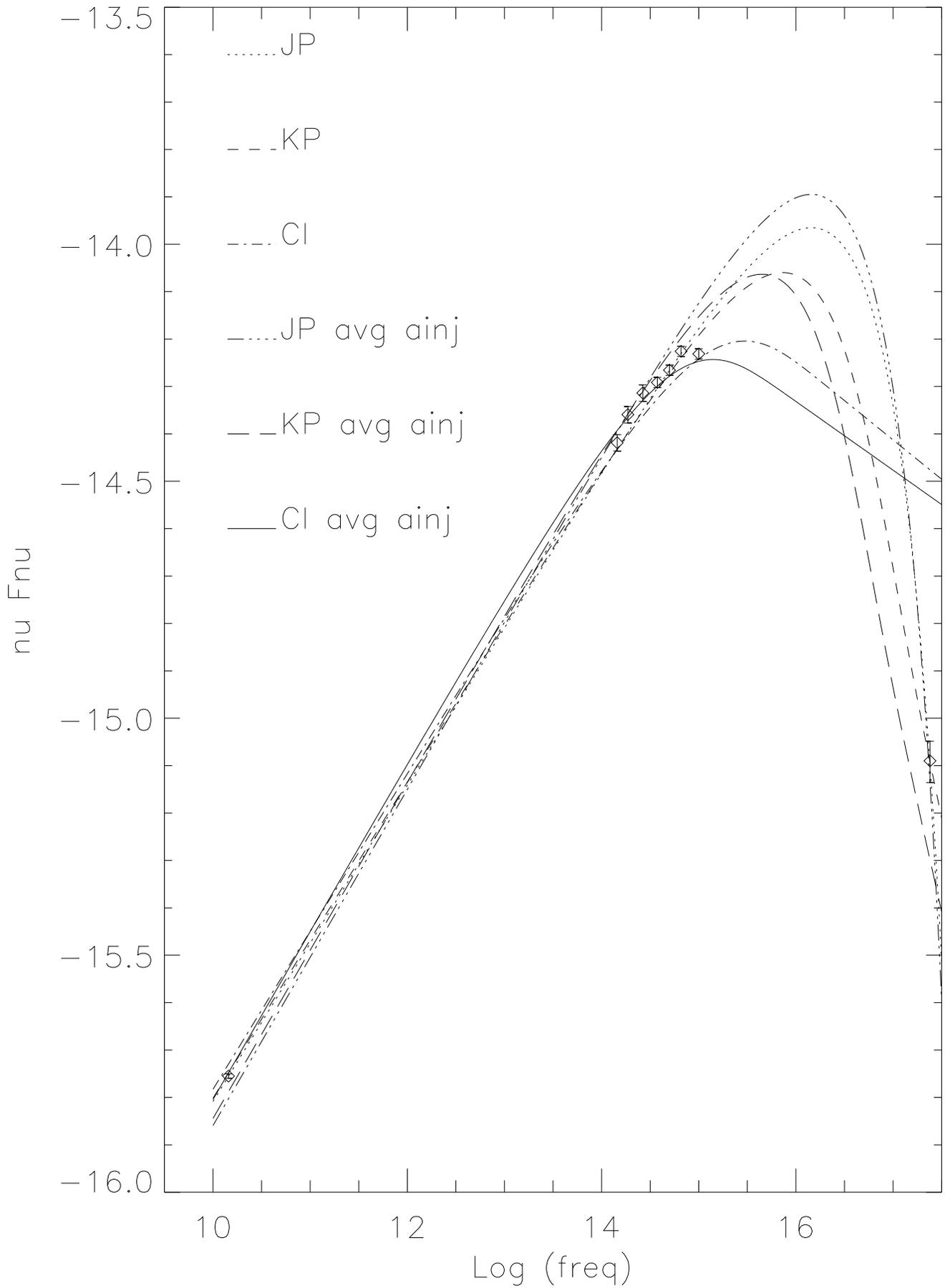
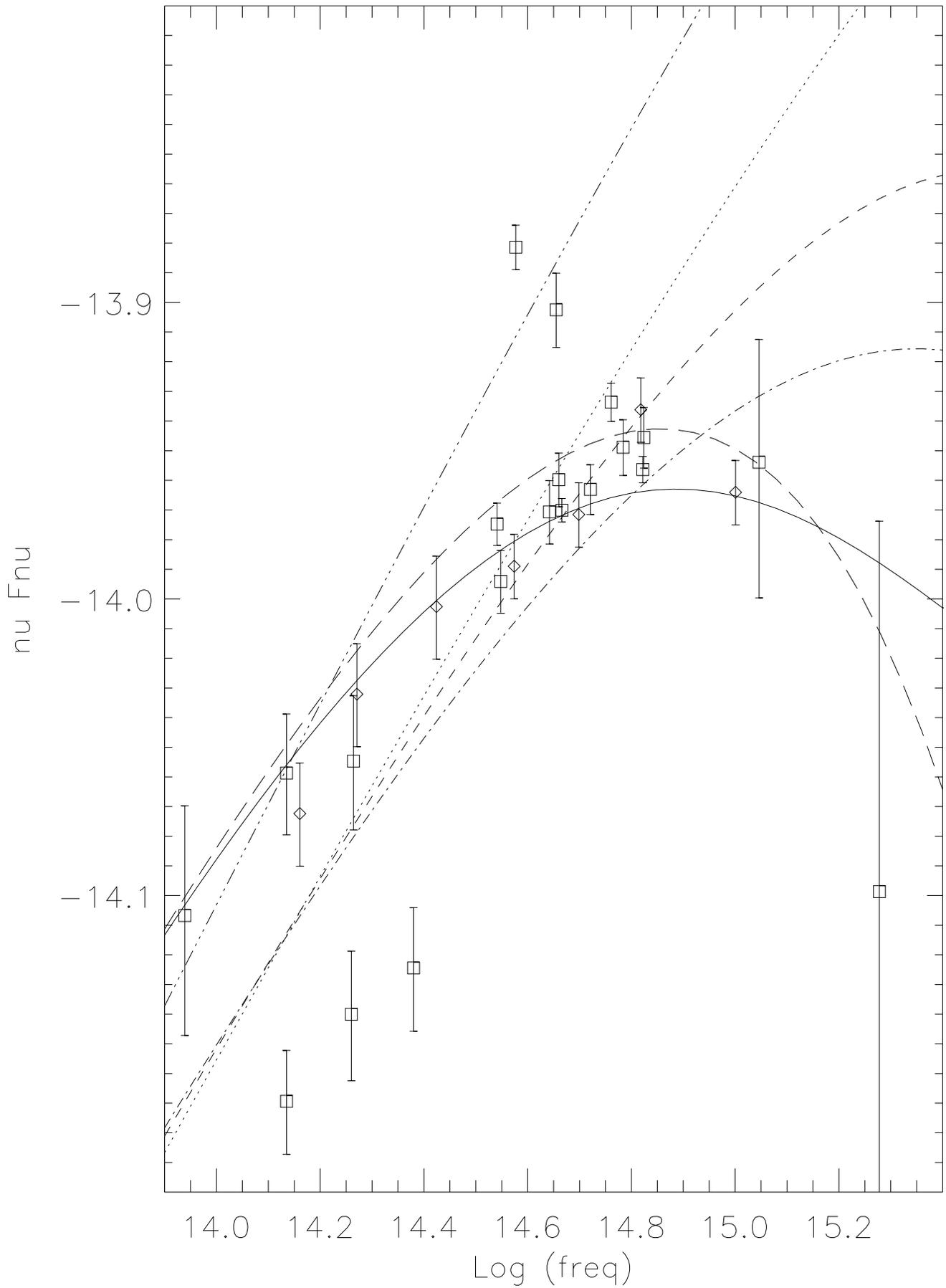
\end{document}